\documentclass{article}
\usepackage{amssymb}
\usepackage{tabularx}
\usepackage{array}
\usepackage{graphicx}
\usepackage{multirow}
\usepackage{float}
\usepackage{hyperref}
\usepackage[utf8]{inputenc}
\usepackage{url}
\usepackage{mathtools,xspace,graphicx,amsmath,amssymb,xcolor}
\mathtoolsset{centercolon}
\usepackage{varwidth}
\usepackage[mathscr]{euscript}
\usepackage{listings}
\newcommand{\subname}[1]{\ensuremath{\textsc{#1}}\xspace}

\definecolor{highlightcolor}{HTML}{F5F5A4}
\definecolor{highlighttextcolor}{HTML}{000000}
\definecolor{bitcolor}{HTML}{a91616}
\newcommand{\codebox}[1]{%
	\begin{varwidth}{\linewidth}%
		\begin{tabbing}%
			~~~\=\quad\=\quad\=\quad\=\kill 
			#1
		\end{tabbing}%
	\end{varwidth}%
}

\newcommand{\fcodebox}[1]{%
	\framebox{\codebox{#1}}%
}

\begin{document}

\title{The Cryptographic Implications of the LinkedIn Data Breach}
\author{
Aditya Gune \\
\texttt{gunea@oregonstate.edu} \\
\\
Department of Electrical Engineering and Computer Science\\
Oregon State University\\
Corvallis, OR 97333 \\
}
\date{March 19, 2017}

\maketitle
\begin{abstract}
Data security and personal privacy are difficult to maintain in the Internet age. In 2012, professional networking site LinkedIn suffered a breach, compromising the login of over 100 million accounts. The passwords were cracked and sold online, exposing the authentication credentials millions of users. This manuscript dissects the cryptographic failures implicated in the breach, and explores more secure methods of storing passwords.
\end{abstract}

\section{Introduction}
LinkedIn is one of the largest social media and professional networking sites in the world. With over 400 million members as of 2017\cite{linkedin_info}, any data leak risks the personal and professional information of millions of people. In 2012, LinkedIn's servers were breached, exposing the hashed passwords of approximately 117 million accounts over a course of several years.\cite{perez} These passwords were quickly decoded and sold online, indicating that LinkedIn's method of storing passwords was cryptographically insecure.

\subsection{The password problem}
The problem of storing passwords securely has existed for decades. Storing passwords in plaintext is clearly insecure, posing an immediate and uncontainable risk in the event of a breach. Storing encrypted passwords also poses a risk due to its two-way nature -- anything encrypted can also be decrypted; in addition, encryption keys are often stored on the same servers as the data being encrypted, meaning a breach is potentially disastrous.

Most websites opt for storing hashed passwords. The intent of storing hashed passwords is to provide a fail-safe to prevent identity theft in the event of a server breach. The method is intended to provide authentication \textit{and} secrecy, and to prevent adversaries from being able to guess the passwords being hashed. 

\subsection{LinkedIn breach} 
Like most web services, LinkedIn hashed its passwords. The company passed user information through a SHA-1 hash function.\cite{goldman} The original breach, speculated to be through a SQL injection attack, occurred on June 5th 2012, and was reported by a number of news agencies. LinkedIn confirmed the breach the following day. \cite{linkedin_blog1} It was first thought that the breach exposed around 6.5 million passwords. However, in 2016, the full dump of the hack was posted, exposing the accounts of over 117 million users, whose information had been compromised but not made public. \cite{vice} The severity of the breach was compounded by improper use of cryptographic hash functions to conceal the password plaintext. 

\section{Cryptography of Password Hashing}
Password hashing works by taking a password input of variable length, and providing a fixed-length output that seems like a random string. SHA-1, the specific hash function used by LinkedIn, was published in the 1990s. SHA-1 can encode an up to $2^{64}$ bit input into a 160 bit message digest.\cite{rfc3174} However, LinkedIn's implementation of SHA-1 failed to include salts -- random numbers unique to every user -- instead simply storing hashed passwords directly to the server.\cite{ars}

SHA-1 works by repeatedly padding the message and breaking it into 512-bit blocks, then iteratively passing these through a series of logical functions. Further information can be found in the official Secure Hash Standard publication FIPS 180-4.\cite{fips180-4} 

\section{Analysis of Failure}
Despite the illusion of secrecy provided by SHA-1, LinkedIn's method of storing passwords was dangerously insecure. Using an unsalted hash leaves the system vulnerable to rainbow table attacks. 
\subsection{Rainbow tables}
Rainbow table attacks assemble a pre-compiled list of passwords into data structures known as \textit{hash chains}. The hashes of these passwords have already been calculated\cite{capec}. However, the storage cost for a simple associative array of passwords to hashes would be very high. Rainbow tables use \textit{reduction functions}, which map individual hashes back to plaintext passwords. This creates a time-memory tradeoff, sacrificing time at the cost of memory.\cite{oechslin} By applying alternating sequences of hashes and reduction functions, an attacker can create chains of hashes and plaintexts:\cite{oechslin}\\
\begin{figure}[H]
\begin{center}
$P_i\xrightarrow{\text{Hash(P)}}C_i\xrightarrow{\text{Red(C)}}P_{i+1} \cdots P_n$
\end{center}

 \caption{A schematic of a hash chain}
 \end{figure}
 Only the first and last elements in the chain -- $P_0$ and $P_n$ -- are stored in memory, thereby greatly reducing the amount of memory required to store a large number of password - hash pairs. Without a salt, an adversary could apply the combination of reduction and hashing functions until they reached an endpoint on a hash chain, then take the corresponding starting plaintext and follow its hash chain until the target password was found.\cite{oechslin} This vulnerability is compounded by the fact that poor password choices are often repeated by multiple users. Security firm KoreLogic's dump of the most common passwords exposed in the LinkedIn attack revealed that over 1 million users used the phrase ``123456" as a password.\cite{korelogic}\\

The introduction of a salt drastically increases the computational cost required for an effective rainbow table. For example, the password `123456' always hashes to the same value, meaning there is only one hash chain entry that needs to be stored for this password. However, if we add a 2-bit salt to the beginning of the password, we see that the adversary now has $2^2$ more possible values, extending the length of the hash chains within the rainbow table:
\begin{lstlisting}
   $ hashlib.sha1(m)
(`5a44cf4f2b0f2bfc7da6f386481f6afbc8aff73f',`01123456')

   $ hashlib.sha1(s + m)
(`5a44cf4f2b0f2bfc7da6f386481f6afbc8aff73f',`01123456')
(`ac0e191df76d3714cb4e2c2659d51753775662d6',`10123456')
(`3cf621ead5cc3885a4a5caef840aad7404bdee81',`11123456')
(`b388959b842429b18180899f7b101cf7ed8667db',`00123456')
\end{lstlisting}

A sufficiently large salt (such as 128 bits) would require the adversary to store or compute $2^{128}$ more values for \textit{each} password, making the cost of such a rainbow table prohibitively high.

\section{Potential Solutions}
Of the most common ways to store passwords, LinkedIn chose perhaps the least secure (aside from storing just plaintexts). Nonetheless, storing passwords in a cryptographically secure way remains a difficult task for system administrators, who often resort to using non-cryptographically secure constructions or hash functions. Principally, cryptographic hash functions should account for the \textit{cost} of computing a hashed password, and ensure that cost scales with faster microprocessors, a requirement that traditional hash functions often do not satisfy. UNIX's \textit{crypt} function was first introduced in 1976, and could hash approximately four passwords per second on contemporary hardware. By 1999, this had increased to 200,000 \textit{crypt} operations per second. \cite{bcrypt} Using GPU-driven parallel computing, password crackers were able to compute over 63 billion SHA-1 hashes per second in 2012. \cite{passwords12} This trend underscores a fundamental requirement of cryptographic hash functions: they must be resistant to brute-force attacks. 

To accomplish this, a password hashing function must ensure that the computational cost of hashing a password is large enough that an adversary attempting to crack thousands or millions of passwords through brute force will find the exercise too time-consuming. 

\subsection{Key Derivation Functions}
A key derivation function (KDF) takes a non-random source of input - such as a password - and derives a cryptographically strong secret key from it. Many contemporary systems for securing passwords use KDFs rather than standard hash functions. A secure KDF will output a key that is indistinguishable from a random string of the same length, and will draw from a uniform distribution of such strings. Most KDFs are constructed from two modules: one that "extracts" a pseudorandom key $K$ from the source input, and one that uses $K$ as the seed for a pseudorandom function (PRF) to produce several cryptographically secure pseudorandom keys.\cite{krawczyk}

However, a KDF in itself is not a cryptographically secure way to store passwords, and may not be resistant to brute-force attacks. A KDF construction used for password hashing should increase the cost required to compute a password hash. The implementations discussed here -- Bcrypt and scrypt -- accomplish this through a method known as \textit{key stretching}. Key stretching involves using a KDF that requires $2^k$ cryptographic operations per hash computation, regardless of the length of the password. For example, an $n$-bit password would require $2^{k+n}$ operations under this system. \cite{scrypt}
\subsection{Bcrypt}
Bcrypt is an adaptive key derivation function, and is perhaps one of the most widely recommended methods for storing passwords securely. As discussed above, password hashing functions must be resistant to brute force attacks. Bcrypt provides this resistance by allowing administrators to specify the number of internal iterations required for a password hash calculation. This causes the computational cost of hashing a password, making brute force attacks prohibitively expensive.
Bcrypt is based on the Blowfish cipher, a ``64-bit block cipher structured as a 16-round Feistel network."\cite{bcrypt} 
\[
\fcodebox{
			\underline{$\subname{Blowfish}(m)$:\cite{bcrypt}} \\
			\>$L_0 := m[:half]$\\
            \>$R_0 := m[half:]$\\
            \>for $i = 1$ to 16:\\
            \>\>$R_i := L_{i-1} \oplus P_i$\\
            \>\>$L_i := R_{i-1} \oplus F(R_i)$\\
            \>$R_{17} := L_{16} \oplus P_{16}$\\
            \>$L_{17} := R_{16} \oplus P_{18}$\\
			\>return $L_{17} || R_{17}$
		}
\]

Though reduced round Blowfish does have vulnerabilities, the standard Blowfish block cipher has been demonstrated secure. \cite{alabaichi}
The Bcrypt algorithm uses the variant \textit{Eksblowfish}, which takes in the user's desired cost of computation as a parameter:\\
\[
\fcodebox{
			\underline{$\subname{EksBlowfish}(cost,salt,key)$:\cite{bcrypt}} \\
			\>$state \gets $ InitState()\\
            \>$state \gets $ ExpandKey(\textit{state, salt, key})\\
            \>repeat $2^{cost}$:\\
            \>\>$state \gets $ ExpandKey(\textit{state, 0, salt})\\
            \>\>$state \gets $ ExpandKey(\textit{state, 0, key})\\
			\>return $state$
		}
\]

Bcrypt is designed to remain resistant to brute force attacks regardless of the speed of modern microprocessors. The ability of Bcrypt to iterate for $2^{cost}$ cycles allows administrators to increase the number of iterations to compensate for stronger password cracking hardware. As the number of iterations increases, individual password hashes take longer to compute, ensuring that a brute force attack remains an expensive task for adversaries.  \cite{malvoni}

\subsection{Scrypt}
Scrypt follows a similar paradigm to Bcrypt, using an adaptive cost mechanism to stymie brute force attackers. However, Bcrypt is still vulnerable to brute force attacks using parallelized hardware. Improvements in semiconductor technology allow attackers to embed more circuits at lower cost. \cite{scrypt} 

Highly parallelized hardware such as the GPU is well suited to the repetitive task of password cracking, because the large number of circuits allows even computationally expensive passwords to be hashed non-sequentially. Billions of password hashing operations can be performed every second, greatly reducing the effectiveness of even adaptive KDFs like Bcrypt. \cite{passwords12}

Scrypt provides resistance against parallelized brute force attacks by making both sides of the time-memory tradeoff costly. It allows the amount of memory required to hash a password to increase proportionally to the number of computations.

\[
\fcodebox{
			\underline{$\subname{MFcrypt}(password,salt,N, p, dkLen)$:\cite{scrypt}} \\
			\>$(B_0 \cdots B_{p-1}) \gets PBKDF2(P, S, 1, p*MFLen)$\\
            \>for $i=0$ to $p-1$:\\
            \>\>$B_i \gets MF(B_i, N)$\\
			\>$DK \gets PBKDF2_{PRF}(P, B_0 || B_1 || \cdots || B_{p-1}, 1, dkLen)$
		}
\]

The full proof of this algorithm can be found in \cite{scrypt}. Scrypt's large memory requirement is partly due to the action of computing and storing in memory a large number of bit strings using a pseudorandom function. This requires allocating a large number of memory locations. The amount of memory used and number of computations required can both be tuned using the parameters $N$ and $p$. This allows system administrators to ensure that the cost of hashing millions of passwords remains very expensive (both in time and in memory), even with advancements in hardware. 

\section{Conclusion}
The use of cryptographically secure password storage remains critical to information security as a whole. Password hashing is a last line of defense, protecting user data in the event of a breach. The 2012 LinkedIn breach is an excellent case study in securing stored passwords. 

The use of an unsalted hash function led to the exposure of over 100 million passwords. As we have shown, standard hash functions such as SHA-1 are no longer adequate for storing passwords, and are vulnerable to both rainbow table and brute force attacks. Alternatives such as Bcrypt and scrypt use combinations of salts and adaptive computation and memory cost, to make password cracking more expensive than what a standard unsalted hash function would allow.

\bibliography{cryptography-linkedin-gune-2017}
\bibliographystyle{acm}

\end{document}